\documentstyle[seceq,supplement,psfig,wrapft]{ptptex}

\title{
Reaction dynamics for fusion of weakly-bound nuclei
}

\author{
Kouichi {\sc Hagino}$^1$ 
and Andrea {\sc Vitturi}$^2$
}

\inst{
$^1$Yukawa Institute for Theoretical Physics, 
Kyoto University, \\ Kyoto 606-8502, Japan \\
$^2$Dipartimento di Fisica, Universita di Padova and INFN, I-35131
Padova, Italy
}

\recdate{
}

\abst{
We discuss several open problems of fusion reactions induced by 
weakly bound nuclei. For this purpose, we solve a one dimensional
three-body Hamiltonian with the coupled-channels formalism. We show
that the continuum-continuum couplings substantially reduce the total fusion
probability at energies above the barrier compared with the no-breakup
case, although the fusion probability 
remains enhanced at subbarrier energies. 
We then discuss a role of transfer process
in fusion of weakly bound nuclei, and point out that  
removing spurious Pauli forbidden transfer components from 
the calculation may be crucial 
at energies below the barrier. 
Calculations based on the three-body classical trajectory Monte Carlo
(CTMC) method are also presented in order to discuss how to 
model complete fusion process. 
}

\begin{document}

\maketitle

\section{Introduction}

The recent availability of radioactive beams has opened up the opportunity 
to study the interactions and structure of exotic nuclei close to the 
drip lines. Unstable neutron-rich nuclei having very weakly bound 
neutrons exhibit characteristic features such as a neutron halo and a 
low energy threshold for breakup. 
These features may dramatically affect fusion and other reaction processes. 

One of the main issues in the study of reaction induced by weakly-bound nuclei
is to clarify how the breakup process influences fusion reactions 
at energies close to the Coulomb barrier.
\cite{HPCD92,TKS93,DV94,HVDL00,DTT02,Y97,YUN03,D99,P02,S98}
In Ref. \citen{HVDL00}, we have performed coupled-channels calculations 
in order to address this question by discretizing in energy the
particle continuum states of the projectile nucleus.  
Defining the complete fusion as
absorption from bound state channels, we have demonstrated that fusion 
cross sections are determined by the competition of two mechanisms, 
i.e., dynamical modulation of fusion barrier and flux loss due to the
breakup couplings. 
Their net effect differs
depending on the bombarding energy: the {\it complete} fusion cross
sections are enhanced at energies below the barrier, while they are
hindered above the barrier, compared with cross sections for a
tightly bound system. On the other hand, 
the {\it total} fusion cross sections (a sum of
complete and incomplete fusion cross sections) are enhanced at 
subbarrier energies while they are close to the no-breakup
calculations at above. 
In these calculations, we did not include the
couplings among the continuum channels. 

In this contribution, we would like to 
discuss how these conclusions may be altered 
by more refined calculations. We particularly pay attention to the 
definition of complete fusion, the role of continuum-continuum
couplings, and the effect of transfer channels. 
To this end, we solve a simple one
dimensional three-body Hamiltonian with the coupled-channels method. 
As we will show below, our main 
conclusions for the complete fusion remain valid at least in a
qualitative way, although we find that the total fusion 
cross sections are hindered at energies above the barrier due to 
the continuum-continuum couplings. 

\section{Separation of complete and incomplete fusion}

\begin{wrapfigure}{r}{6.6cm}
\vspace*{0.2cm}
    \parbox{6.6cm}
           {\psfig{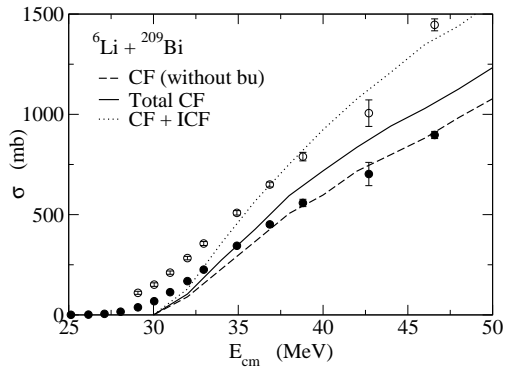}}
\protect\caption{
Comparison of the three-body classical trajectory Monte Carlo
calculations with the experimental data for 
$^{6}$Li +$^{209}$Bi reaction. 
The solid and the dotted lines are complete and total fusion cross
sections, respectively. The contribution of the absorption of the
whole projectile from bound states is denoted separately by the dashed
line. 
The experimental data are taken from Ref. \cite{D99}. }
\end{wrapfigure}

Theoretically, the complete fusion refers to the capture of the
whole projectile, while the incomplete fusion is a process where only
a part of the projectile is captured. 
In our previous calculations \cite{HVDL00}, 
we defined complete fusion as absorption from bound state channels,
while the incomplete fusion as those from continuum state channels in
the projectile. The basic idea was that a continuum state is a
doorway of breakup, and thus is related to incomplete fusion process. 
However, when the breakup takes place inside the Coulomb 
barrier, all the breakup fragments would likely be captured by the
target nucleus, leading to complete fusion reaction.\cite{D99,P02} 
This breakup
followed by complete fusion was not taken into account in the
definition of complete fusion in Ref. \citen{HVDL00}, and  
the previous calculation provides only the lower limit of complete
fusion cross sections. 

In order to estimate the contribution of the breakup followed by
complete fusion, one of us has developed a three body classical
trajectory Monte Carlo (CTMC) method. \cite{D99} 
For a three body Hamiltonian that consists of the target ($T$) and two
projectile fragments ($P_1$ and $P_2$), two-dimensional classical
Newtonian equations are solved to obtain the time evolution of the 
coordinates and velocities of the fragments. The initial conditions
are that the projectile, with its two fragments in random orientation, 
starts far from the target with impact parameter $b$. As the
projectile moves towards the target, three processes are possible,
depending the value of the impact parameter $b$: (i) the projectile as
a whole or both of the fragments are absorbed by the target, (ii) only
one fragment is absorbed, and (iii) neither fragment is captured. 
We assume that the breakup occurs when the distance between the
projectile fragments exceeds their potential barrier radius, 
and that the projectile fragment $P_i$ 
is absorbed by the target nucleus when the
relative distance between the target and the fragment 
is smaller than $r_{\rm abs}=1.1\times (A_T^{1/3}+A_{P_i}^{1/3}$). 
See also Ref. \citen{HDH03} for details. 

Figure 1 shows the results of the three body CTMC method for the fusion 
reaction of $^{6}$Li +$^{209}$Bi. \cite{D99} 
The dotted line denotes the sum of complete and incomplete fusion 
cross sections. 
Since this is a classical calculation, cross sections are finite 
only at energies above the Coulomb barrier. 
The dashed line indicates the absorption of the whole projectile from the
bound states, while the solid line shows the total complete fusion cross
sections. The difference between the solid and the dashed line
thus represents the contribution from the breakup followed by 
the complete fusion process. 
As we see, there is a significant
contribution of this process at energies above the barrier. 
This process has not been considered in any coupled-channels
calculations so far, and the present result suggests that 
a consistent definition of complete fusion is necessary when one
compares experimental data with theoretical calculations. 

\section{Role of continuum-continuum couplings}

Let us now discuss the effect of continuum-continuum couplings in the 
coupled-channels calculations. 
Diaz-Torres and Thompson showed that this effect increases the
irreversibility of breakup process, and thus hinders the total fusion
cross sections\cite{DTT02}. 
In fact, their calculation suggests that the total fusion cross sections are
smaller than the no-coupling case at energies above the barrier,
although those are larger at energies below the barrier. 
A similar hindrance has been found in 
the time dependent wave packet approach of 
Yabana {\it et al.} \cite{Y97,YUN03}. 
Their calculations indicate that the total fusion probability is 
smaller than that for the core-target system, suggesting that it is 
much smaller than the prediction of the folding model calculation, 
which statically takes into account the halo structure of the
projectile nucleus. They attribute this hindrance of fusion
probability to a spectator role of the valence 
neutron\cite{Y97,YUN03}. 
In this section, we revisit this problem and discuss whether the hindrance of
total fusion probability observed in the time dependent wave packet
approach can be understood in terms of the effect of continuum-continuum
couplings in the coupled-channels approach. 

To this end, we use a one dimensional three-body Hamiltonian which 
was used by Yabana and Suzuki for their time-dependent wave packet 
calculation for fusion of $^{10}$Li + $^{40}$Ca system\cite{Y97}. 
Assuming that the projectile consists of a core nucleus ($C$) and a
valence neutron ($n$), and that the target mass is infinite, the model
Hamiltonian reads \cite{Y97}
\begin{equation}
H=
-\frac{\hbar^2}{2M}\frac{\partial^2}{\partial X^2}+V_{CT}(X-\frac{m_n}{M}x)
+V_{nT}(X+\frac{m_C}{M}x)
-\frac{\hbar^2}{2\mu}\frac{\partial^2}{\partial x^2}
+V_{nC}(x)
\end{equation}
where $X$ and $x$ are 
the relative coordinate between the target and the center of mass of
the projectile, and that between the core and the valence neutron in the
projectile, respectively. 
$M=m_n+m_C$ is the total mass of the projectile, while $\mu=m_n\cdot m_C/M$ 
is the reduced mass for core-neutron system. 
We use the same potentials with the same values for the parameters as
in Ref.\citen{Y97}, except for the $X_M$ parameter in the Coulomb
interaction which we somewhat increase in order to increase the
barrier height.

\begin{wrapfigure}{r}{6.6cm}
\vspace*{1cm}
    \parbox{6.6cm}
           {\psfig{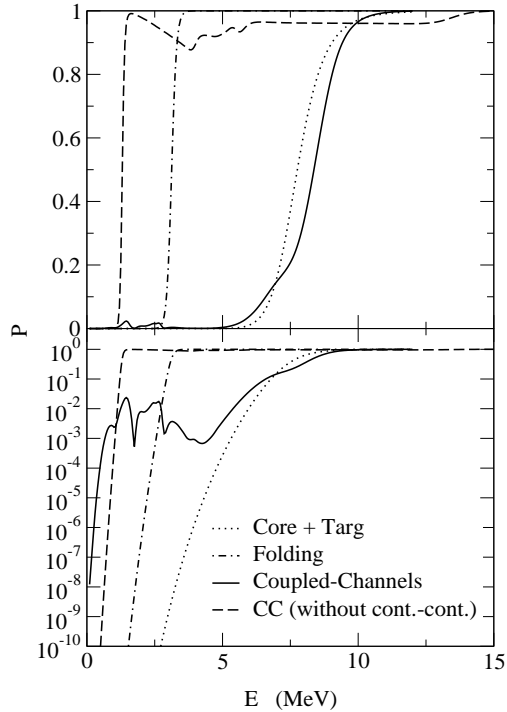}}
\protect\caption{
The total penetrability for the one dimensional three body Hamiltonian 
obtained by the several methods. The solid and the dashed lines are
the results of the coupled-channels calculations with and without the
continuum-continuum couplings, respectively. The dot-dashed line is
the penetrability for the folding potential, while the dotted line
denotes the penetrability for the core-taget system. }
\end{wrapfigure}

In order to solve this Hamiltonian with the coupled-channels
framework, we first obtain the eigenstates for the projectile 
Hamiltonian, $H_{\rm proj}=-\frac{\hbar^2}{2\mu}\frac{d^2}{d x^2}
+V_{nC}(x)$, by expanding the wave functions on the orthogonal
polynomial basis, which was 
recently advocated by 
P\'erez-Bernal {\it et al.}\cite{PBMAGC03}. 
This basis is constructed from the 
ground state wave function of the Hamiltonian $\phi_0(x)$ by
multiplying a polynomial function to it 
($\phi_0, x\phi_0, x^2\phi_0, \cdots, x^n\phi_0$). 
Since the wave function is expanded on the discrete basis, 
the continuum states, obtained by diagonalizing the Hamiltonian,  
are also discretized in energy. We use the polynomial
basis up to $n$=28. The energy for the first five states are
given by: $\epsilon_0=-26.09$ MeV,  $\epsilon_1=-11.42$ MeV,  
$\epsilon_2=-0.515$ MeV,  $\epsilon_3=0.135$ MeV, and 
$\epsilon_4=0.193$ MeV. Following Ref. \citen{Y97}, we assume that 
the ground and the first excited states in this Hamiltonian 
are occupied by the 
neutrons in the core nucleus, and consider the second 
excited state as the ground state of the core-neutron system, $\phi_{gs}(x)$. 

Figure 2 shows the total barrier penetrability obtained with 
the coupled-channels method 
(the solid line) in the
linear (the upper panel) and in the logarithmic scale (the lower
panel). We include the first seven 
discretized continuum states together with the ground state. 
At low energies, some of the continuum states are kinematically
forbidden (i.e., closed channels), and 
it is not easy to obtain numerically stable solutions of 
the coupled-channels equations. 
We avoid this problem by using the modified log-derivative method.\cite{BB94}  
This method is in fact very powerful, and we have confirmed that our
results are stable against the mesh size and the matching radius. We
have also checked that the low energy penetrability does not change
much even when we include more continuum state channels in the
coupled-channels equations. The penetrability still shows a
resonance-like structure. This occurs at a threshold energy for each
discretized channel, and can thus be attributed to the threshold
effect. \cite{HB04}

The result of the coupled-channels calculations is compared with 
the fusion probability (barrier penetrability) 
for the folding potential (the dot-dashed line), 
\begin{equation}
V_{\rm fold}(X)=\int\,dx \phi_{gs}(x)^2
\left[V_{CT}(X-\frac{m_n}{M}x)
+V_{nT}(X+\frac{m_C}{M}x)\right].
\label{folding} 
\end{equation}
We find that the total penetrability is significantly hindered by the 
continuum couplings at energies above the barrier height of the
folding potential, although it is still enhanced compared with the
penetrability for the folding potential at energies below the
barrier (see the lower panel). Notice that the total penetrability 
is even smaller than 
the penetrability for the core-target system 
(the dotted line) at energies above the barrier. 
This is reminiscent of the result of the time dependent wave packet
approach. \cite{Y97}

The result without the continuum-continuum couplings is denoted by the
dashed line. As was pointed out in Ref. \citen{DTT02}, this
calculation significantly overestimates the barrier penetrability
compared with the full coupled-channels calculation. 
A part of the reason why the continuum-continuum couplings reduce the
penetrability can be understood 
as follows: the wave function of the (discretized) 
continuum state considerably extends outside the barrier. The diagonal 
matrix element of the coupling Hamiltonian for a continuum state channel 
is thus much smaller than 
that for the ground state, eq. (\ref{folding}). 
Notice that the
diagonal matrix element is given by the folding potential itself 
for all the channels if 
the continuum-continuum couplings are neglected. 
The change of the diagonal matrix elements can be regarded as
``effective Q-value'', which is large and positive around the
barrier region. If the coupling is strong enough, 
the penetrability can then be reduced due to the so called
anti-adiabatic effect. \cite{RTN92,OK96} 

\section{Role of transfer process}

We next discuss the effect of transfer couplings on fusion of a halo
nucleus. Compared with the breakup effects, a discussion on the
transfer effect has yet been scarce. \cite{OK96}
The time dependent wave packet 
approach of Yabana {\it et al.} shows that the fusion probability 
is hindered over that for the core-target system even at energies 
well below the
barrier. \cite{YUN03} This conclusion is qualitatively different from
the result of the coupled-channels approach,\cite{HVDL00,DTT02} 
which predicts an
enhancement of fusion probability at low energies, as was discussed in the
previous section. 
In both of these approaches, the multipole 
decomposition is usually introduced to the total wave function. 
Yabana {\it et al.} examined the effect of the angular momentum
truncation for the neutron-core system, and showed that the fusion 
probability is enhanced when the angular momentum is limited to low
values.\cite{YUN03} A high value for the angular momentum 
is required in order to simulate the transfer
process of the neutron to the target nucleus, and 
Yabana {\it et al.} claimed that the enhancement of fusion cross
sections found in the coupled-channels approach is an artifact of
neglecting the transfer coupling. 
They needed up to $l=70$ in order to get a converged result. 
Diaz-Torres and Thompson also studied the effect of angular momentum
truncation in the coupled-channels framework. \cite{DTT02}
Their result seems to indicate that the fusion cross sections indeed get
smaller when higher anuglar momentum components are included in the
calculation (see Fig. 6 in Ref. \citen{DTT02}). 

\begin{wrapfigure}{r}{6.6cm}
\vspace*{1.cm}
    \parbox{6.6cm}
           {\psfig{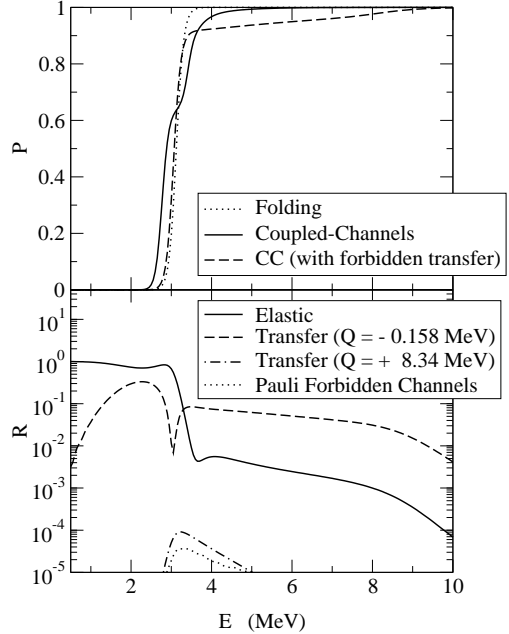}}
\protect\caption{
The effect of the Pauli forbidden transfer channels on the
transmission (the upper panel) and the reflection (the lower panel)
probabilities.  
The dashed and the solid lines in the upper panel are obtained by
solving the coupled-channels equations with and without the Pauli
forbidden channels, respectively. The dotted line in the upper panel 
is the penetrability 
for the folding potential. 
The solid line in the lower panel is the reflection probability for
the incident channel obtained by the coupled-channels calculation with
the Pauli forbidden channels. The dashed and the dot-dashed lines are
the transfer probabilities for the highest and the second highest
bound states in the target nucleus. The total reflection probability for the
Pauli forbidden channels is denoted by the dotted line. }
\end{wrapfigure}

However, we here ask a general question: 
is it really physical to include all the partial wave 
components for the neutron-core system? 
It is true that 
one can in principle solve the three-body Hamiltonian exactly 
if one includes all the partial waves. 
In such a calculation, {\it all} the transfer channels are automatically 
entered. 
Notice that these include even a transfer to a deep bound
state of the target nucleus, which must be forbidden by the Pauli
principle. 
Those Pauli forbidden transfer channels most likely
have a large positive $Q$ value, and the fusion cross sections may be
artificially hindered due to the anti-adiabatic effect \cite{RTN92} if
these channels are included in the calculation. 
Also, when the initial configuration of the valence neutron
is at an excited state of the Hamiltonian 
due to the Pauli principle, the forbidden
de-excitations to lower states in the projectile may be entered 
unless those transitions are explicitly excluded by {\it e.g.} the
basis expansion. Those forbidden de-excitations may also lead to
an artificial reduction of fusion cross sections for the same reason. 

In order to examine the effect of the spurious Pauli forbidden
transfer channels in the
projectile, we again solve the one dimensional three body model in the
previous section with the coupled-channels method by including several 
transfer channels. For the parameter set specified in the previous
section, there are five bound states in the target nucleus at 
$-40.5, -31.9, -20.6, -8.85$ and $-0.357$ MeV. 
We assume that the ground state as well as the first and the second excited
states in this Hamiltonian are already occupied. The solid line in the
upper panel of Fig. 3 is the result of the coupled-channels
calculation where only the transfer channels to the two allowed states 
in the target are included. We compute the transfer coupling matrix
elements in the no-recoil approximation using the neutron-target
potential $V_{nT}$ as the transfer interaction. In order to mimic the
recoil effect, we multiply a factor of $2(m_C+1)/(2m_C+1)$ to the
transfer matrix elements.\cite{EFL89,QPBW85} 
For simplicity, the continuum excited states in the projectile are
not included. In this calculation, 
one can clearly see that the fusion 
probability is enhanced due to the transfer couplings at energies 
below the barrier compared with the folding model calculation (the
dotted line). The result of the coupled-channels calculation with the
Pauli forbidden channels is given by the dashed line. 
One sees that the enhancement of 
fusion probability is much less at energies below the barrier 
due to the anti-adiabatic effect. 

The lower panel of Fig. 3 shows the reflection 
probabilities for the elastic (the solid line) and the transfer
channels. The dashed and the dot-dashed lines represent the transfer
probability for the highest and the second highest bound states in the
target nucleus,
respectively. Due to the $Q$-value matching, the total transfer probability
is dominated by the former, as has been found in the time dependent
wave packet approach. \cite{Y97} 
The dotted line is the total reflection probability for the Pauli
forbidden channels. Notice that this probability is much smaller than 
the transfer probability for the allowed channels, and may be
negligible compared to the latter. 
Yet, the Pauli forbidden processes influence the barrier 
penetrability (the fusion probability) in a significant way. 
We thus conclude that care must be taken in discussing the angular
momentum truncation, especially at energies around the barrier. 

\section{Summary and discussion}

Two alternative approaches have been used in the literature 
in order to discuss the subbarrier fusion reaction induced by a 
weakly bound nuclei. A traditional approach is to solve the coupled-channels 
equations by discretizing in energy 
the continuum states in the projectile nucleus (CDCC). 
The second approach 
employs the time dependent dynamics and solves the time 
evolution of a wave packet. 
If one neglects the dynamical effects, the total fusion probabilities 
would have been larger 
than that for the core-target system, 
since the barrier height is considerably reduced due to 
a large extension of neutron wave function in the projectile nucleus. 
On the contrary, the time dependent approach 
predicts smaller fusion probabilities, suggesting 
that there is 
a huge dynamical effect which compensates the static effect. 
Using a simple one dimensional three body Hamiltonian, we have demonstrated 
that this result can be understood in terms of the effect of couplings among 
the continuum states. 
In this conference, Signorini {\it et al.} reported that the measured fusion 
cross sections for the $^{11}$Be+$^{209}$Bi are similar to those 
for the $^{10}$Be+$^{209}$Bi system despite the halo structure of the 
$^{11}$Be nucleus.\cite{S98} 
It might be that the enhancement of fusion cross sections 
due to the static effect (reduction of the barrier height due to 
the halo structure) is compensated by the dynamical breakup effect, 
resulting in similar cross sections between the two systems. 

At energies below the barrier, the two theoretical approaches 
predict a contradictory result to each other. 
In the coupled-channels approach, 
the fusion probability is found to be larger 
than the prediction of the folding model. 
In contrast, 
the time dependent wave packet approach predicts a hindrance 
of fusion probability even 
at energies below the barrier. 
We have argued that care must be taken in the partial wave
decomposition for the wave functions 
because spurious Pauli forbidden transfer channels are 
inevitably coupled when all the partial waves are included. 
We have found that the Pauli forbidden 
transfer couplings significantly hinder the total fusion probability 
even if the transfer probability itself may be negligible. 

In order to draw a more definite conclusion on fusion cross sections 
for a weakly bound system at energies below the barrier, 
further calculations would be required by including 
only the Pauli allowed transfer channels together with the breakup
channels. 
The time dependent wave packet approach would need to introduce 
the projection operator in order to exclude explicitly the Pauli forbidden 
channels. In the coupled-channels approach, the angular momentum 
between the core and the valence neutron should be limited to low values 
in order to avoid the Pauli forbidden transfer, and should explicitly include 
the transfer couplings in addition to the breakup channels. The combined 
effect of breakup and transfer in the coupled-channels approach 
has not been studied so far in the context of subbarrier fusion, 
and it would be an interesting future work. 

A further 
complication arises from a separation of complete and incomplete fusion 
cross sections. Using the three body classical trajectory Monte Carlo method, 
we have demonstrated that there is a significant contribution of the breakup 
followed by complete fusion process to the total complete fusion cross 
sections. It would be another interesting future problem to model 
this process using a quantum mechanical model for subbarrier fusion of 
unstable nuclei.

\end{document}